# The limit behavior of the evolution of Tsallis entropy in self-gravitating systems

Yahui Zheng[a,b], Jiulin Du[b], Faku Liang[a]

*a: Department of Physics, School of Science, Qiqihar University, Qiqihar City 161006, China.*
*b: Department of Physics, School of Science, Tianjin University, Tianjin 300072, China.*

**Abstract** In this letter, we study the limit behavior of the evolution of Tsallis entropy in self-gravitating systems. The study is carried out under two different situations, drawing the same conclusion. No matter in the energy transfer process or in the mass transfer process inside the system, when nonextensive parameter $q$ is more than unity, the total entropy is bounded; on the contrary, when this parameter is less than unity, the total entropy is unbounded. There are proofs in both theory and observation that the $q$ is always more than unity. So the Tsallis entropy in self-gravitating system generally exhibits a bounded property. This indicates the existence of global maximum of Tsallis entropy. It is possible for self-gravitating systems to evolve to thermodynamically stable states.

**1. Introduction**

In the classical Boltzmann-Gibbs (BG) statistics, the Boltzmann entropy of a self-gravitating system is proved to be unbounded [1-3]. This means that in any limited self-gravitating system the entropy can increase infinitely. Such a system can not be at a stable equilibrium state as the increase principle of entropy would compel the system to evolve to a state with higher entropy. In this process the system always exhibits instability, such as the Antonov instability [1] or gravothermal catastrophe [2], which is characterized by the negative heat capacity.

If the classical BG statistics is valid in the self-gravitating systems, the instability should be ubiquitous in the universe. This seems to indicate that most of the celestial bodies can not exist for long term, due to which we will find nowadays so many remains as the results of astrophysical revolution. In other words, we will see a lot of black holes with various mass (even with very small mass), neutron stars or white dwarfs in our universe. This is of course not closer to the fact. Up to now we don't find any report that there are many small black holds, or neutron stars, nearby a observable star, say, our sun. In order to explain this obvious conflict, we have to try to accept other ideas such as that the classical statistics might not be valid in the self-gravitating systems.

Indeed, the fundamental assumption of the BG statistics is that the interaction between particles can be ignored or has the nature of short range. The short range interaction guarantees the additivity of thermodynamic quantities such as energy and entropy. That is to say, in BG statistics, the long range interaction does not participate into the statistical average of particles' kinetic behavior. However, essentially the long range gravitational interaction in self-gravitating systems must enter into the statistical behavior of the particles as a collective. In this sense we can judge that the classical BG statistics is not appropriate in such systems with long range interactions.

It has been accepted universally that the newly proposed nonextensive statistics [4] is suitable for the self-gravitating systems, where the long range interactions play a dominant role. Space observations have showed that the suprathermal halo in astrophysical plasmas is represented by the family of $\kappa$-distributions, which can be explained in theory through the nonextensive statistics [5]. Other space observations have also showed that the turbulence in solar wind manifests a strong non-Gaussian feature and it is argued that this should be related physically to the nonextensive character of the turbulent intermittency of the planetary medium [6]. The similar power law patterns about dark matter and gas density profile are also observed in relaxed galaxies and clusters, which can be easily derived from nonextensive statistics [7]. It has been verified that the rotational velocity distribution of main sequence field stars can not be fitted by the Maxwellian distribution, but can be fitted by the power-law distributions in nonextensive statistics [8].

By the application of nonextensive statistics into self-gravitating systems, it is possible to resolve the conflict between theoretical instability and observations. This issue is closely related to the evolution character of entropy, where the nature of nonextensive parameter $q$ plays important role. There are two aspects about the nature of this parameter: one is its physical origin, and another is its value range.

For the first aspect of nonextensive parameter, the fundamental view is that in different system it has different physical origin. For example, in self-gravitating systems [9-10] the parameter is related to the long range gravity, and also to the gradient of temperature. If the system is rotating, the parameter is further linked to the initial centrifugal force [11]. In plasmas, the parameter is associated to the Lorentz force and the gradient of temperature [12-13]. In stochastic process, this parameter is originated from the inhomogeneity of phase space [14]. This aspect is always related to the fundmental issue, i.e., the definition of temperature; therefore it has attracted more attention.

However, people pay little attention on the second aspect, i.e., the value range the nonextensive parameter possibly locates in. For this point, there are some interesting works in numerical test with astrophysical data. De Freitas and De Medeiros show that the radial velocity distributions of 6781 working samples in the solar neighborhood are best fitted by q–Gaussians within the nonextensive statistics, giving special value of $q$ [15]. They also verify through the study of the rotation distribution of working sample that there are typical values of q about different spectral types, such as with 1.9 for F type and with 2.2 for G type [16].

In this letter, we study the limit behavior of entropy evolution in the representation of nonextensive statistics, where the value range of $q$ plays an important role. For a continuous Hamiltonian system, the Tsallis entropy can be expressed as

$$S_q = -k \int F^q \ln_q F d^{3N}\mathbf{v} d^{3N}\mathbf{r} = -k \int \frac{F - F^q}{1-q} d^{3N}\mathbf{v} d^{3N}\mathbf{r} \quad , \tag{1}$$

where the $N$ is the particle number of the system, and the quantity $F$ is the ensemble distribution function, i.e. Gibbs function of a self-gravitating system, which is composed of gravitationally Hamiltonian particles. It is obviously different from the single-particle distribution function in the generalized physical kinetics [17]. In next sections, we would discuss the evolution of Tsallis entropy both in the energy transfer and the mass transfer processes, confined in an isolated self-gravitating system.

**2. The entropy evolution in the energy transfer**

For our aim let us analyze a self-gravitating system, which have a fixed energy $E$ and a fixed mass $M$. Now we divide the system into two parts: the inner part and outer part. The inner part have mass $M_1$, particle number $N_1$ and radius $r_1$, while the outer part have mass $M_2$ much less than $M_1$, particle number $N_2$ and radius $r_2$. For simplicity, we assume that in each part the particles obey the uniform distribution in phase space (i.e., in each velocity and coordinate space). This means there are up limits of the velocities in each part, which are represented by $v_1$ and $v_2$ respectively.

Although being sufficiently simplified, the above uniform distribution model can not affect the ultimate conclusion in this letter. That is because our aim is to observe the limit evolution behavior of the considered self-gravitating system, under which condition the initial distribution would be quickly forgotten by the evolving system.

In this model, we fix actually the masses of each part, under which the system evolution mainly exhibits the energy transfer from one part to another part. In order to analyze the entropy evolution character in this process, let us write down the uniform distribution in each part, namely,

$$F_1 = \frac{1}{V_1^{N_1}}, \qquad F_2 = \frac{1}{V_2^{N_2}}, \qquad (2)$$

where the phase volumes of single particle are respectively written as

$$V_1 = \frac{16}{9}\pi^2 r_1^3 v_1^3, \qquad V_2 = \frac{16}{9}\pi^2 (r_2^3 - r_1^3) v_2^3. \qquad (3)$$

According to eq. (1), the Tsallis entropies in each part can be calculated by, respectively,

$$S_{q1} = k\frac{V_1^{N_1(1-q)} - 1}{1-q}, \qquad S_{q2} = k\frac{V_2^{N_2(1-q)} - 1}{1-q}. \qquad (4)$$

Then the total Tsallis entropy of the self-gravitating system is given by

$$S_q = S_{q1} + S_{q2} + \frac{1-q}{k}S_{q1}S_{q2} = S_{q1} + S_{q2}(1 + \frac{1-q}{k}S_{q1})$$
$$= k\frac{V_1^{N_1(1-q)} - 1}{1-q} + k\frac{V_2^{N_2(1-q)} - 1}{1-q}V_1^{N_1(1-q)}. \qquad (5)$$

Furthermore, the total energy of the self-gravitating system is [3]

$$E = -\frac{3}{5}G\left[\frac{1}{2}\frac{M_1^2}{r_1} + \frac{5}{3}\frac{M_1 M_2}{r_{12}}\right], \qquad (6)$$

where we have considered the virial theorem and the assumption $M_1 \gg M_2$, and $r_{12}=r_2-r_1$. Now we assume the energy transfers from inner part to outer part, which is actually the usual evolution manner in astrophysical systems. According to the above equation, with the energy loss of inner part, its radius tends to finite value while the radius of outer part tends to infinity. This means that $r_{12} \approx r_2$. According to the virial theorem, there are two equations

$$\frac{3}{10}\frac{GM_1^2}{r_1} = \frac{1}{2}M_1 v_1^2, \qquad \frac{M_1 M_2}{r_2} = \frac{1}{2}M_2 v_2^2. \qquad (7)$$

Therefore, the phase volumes of single particle change into

$$V_1 = \frac{16}{9}\pi^2 (\frac{3}{5}GM_1 r_1)^{\frac{3}{2}}, \qquad V_2 \approx \frac{16}{9}\pi^2 (2GM_1 r_2)^{\frac{3}{2}}. \qquad (8)$$

There is no doubt that in limit of energy transfer, the volume $V_1$ tends to a finite value while the volume $V_2$ tends to infinity.

Now we discuss the evolution behavior of Tsallis entropy in two different value ranges of the nonextensive parameter. In the first case, the nonextgensive parameter is less than unity ($q<1$). According to (5), the total entropy of the self-gravitating system tends to infinity, namely,

$$S_q \to k\frac{V_2^{N_2(1-q)}}{1-q}V_1^{N_1(1-q)} \to +\infty. \tag{9}$$

In the second case, the nonextensive parameter is more than unity ($q>1$). Due to the infinite phase volume $V_2$, the total entropy tends to

$$S_q = k\frac{V_1^{N_1(1-q)}-1}{1-q} - k\frac{1}{1-q}V_1^{N_1(1-q)} = \frac{k}{q-1}. \tag{10}$$

The total Tsallis entropy this time is finite. It can be seen that when $q \to 1$, the above equation tends to infinity, recovering to the classical result [3].

**3. The entropy evolution in the mass transfer**

In this section, we also suppose a self-gravitating system with $N$ particles, still satisfying the uniform distribution. Now divide it into two subsystems. Some particles $\alpha N$ are homogeneously distributed in a spherical space with a radius $r_1$ and they satisfy the same uniform distribution in velocity space, with a up limit of velocity $v_1$. The remained particles $(1-\alpha)N$ are homogeneously distributed in another sphere with radius $r_2$, and satisfy the same distribution in velocity space with a up limit $v_2$. Furthermore, we assume these two subsystems are far from each other, out of touch and with a distance $r_{12}$

Obviously the above uniform distribution model is also sufficient simplified in the strictest sense. Yet now that we only concern the limit evolution behavior of system, this simplified model as an initial distribution does not affect the ultimate conclusion.

In this model, these two subsystems are out of touch each other, implying the main evolution manner is the mass transfer. In order to analyze the entropy evolution in this process, give out the distribution functions of these two subsystems, that is [2],

$$F_1 = \frac{1}{V_1^{\alpha N}}, \qquad F_2 = \frac{1}{V_2^{(1-\alpha)N}}, \tag{11}$$

where the phase volumes of single particle are respectively written as

$$V_1 = \frac{16}{9}\pi^2 r_1^3 v_1^3, \qquad V_2 = \frac{16}{9}\pi^2 r_2^3 v_2^3. \tag{12}$$

The total Tsallis entropy of the self-gravitating system is also given by

$$S_q = S_{q1} + S_{q2} + \frac{1-q}{k}S_{q1}S_{q2} = S_{q1} + S_{q2}(1+\frac{1-q}{k}S_{q1})$$

$$= k\frac{V_1^{\alpha N(1-q)}-1}{1-q} + k\frac{V_2^{(1-\alpha)N(1-q)}-1}{1-q}V_1^{\alpha N(1-q)}. \tag{13}$$

Furthermore, the total energy of the considered self-gravitating system is [2]

$$E = -\frac{3}{5}GM^2\left[\frac{\alpha^2}{r_1} + \frac{(1-\alpha)^2}{r_2} + \frac{5}{3}\frac{\alpha(1-\alpha)}{r_{12}}\right] + \frac{3}{10}M[\alpha v_1^2 + (1-\alpha)v_2^2]. \tag{14}$$

The quantity $M=Nm$ is the total mass of the system and $m$ is the mass of single particle. Now we assume that the mass transfers from subsystem 2 to subsystem 1 due to some unknown reasons. In this process, the virial theorem is impossible to hold. Therefore we need to add an artificial restriction to the mass transfer process. Actually, we can directly adopt the treatment method in [2]. That is, we analyze the change of total Tsallis entropy, in the limits $\alpha \to 1$ and $r_2 \to 0$, to keep the following expression unchanged in mass transfer, namely

$$(1-\alpha)\ln(\tfrac{16}{9}\pi^2 r_2^3 v_2^3) = -L, \tag{15}$$

where $L$ is a constant. Notice that there are two variables in each phase space. For simplicity, we fix $r_1$ and let $v_1$ change in subsystem 1; similarly, we fix $v_2$ and let $r_2$ change in subsystem 2.

In order to determine the tendency of $v_1$ in the limits $\alpha \to 1$ and $r_2 \to 0$, we rewrite (14) as

$$v_1^2 = \frac{1}{\alpha}\left[\frac{10}{3M}\left\{E + \frac{G\alpha(1-\alpha)M^2}{r_{12}}\right\} + \frac{2G\alpha^2 M}{r_1} - (1-\alpha)v_2^2 + \frac{2GM(1-\alpha)^2}{r_2}\right]. \tag{16}$$

The last term in this solution is what we concern. In the limits $\alpha \to 1$ and $r_2 \to 0$, we notice that equation (15) is simplified as

$$3(1-\alpha)\ln r_2 = -L. \tag{17}$$

According to above equation, the last term in (16) can be transformed into

$$\frac{2GM(1-\alpha)^2}{r_2} = \frac{2GML^2}{9r_2(\ln r_2)^2}. \tag{18}$$

It is apparent that in the limit $r_2 \to 0$, the above expression tends to positive infinity. Therefore according to (16) there is $v_1 \to +\infty$, leading to $V_1 \to +\infty$.

Similarly, we discuss the limit behavior of entropy evolution in two different cases. In the first case, the nonextensive parameter is less than unity ($q<1$). According to (13), in limits $V_1 \to +\infty$ and $\alpha \to 1$ the total Tsallis entropy tends to

$$S_q \to k\frac{V_1^{N(1-q)}}{1-q}e^{-(1-q)LN} \to +\infty. \tag{19}$$

In the second case, the nonextensive parameter is more than unity, i.e., $q>1$. So in the limits $V_1 \to +\infty$ and $\alpha \to 1$ the total entropy tends to

$$S_q \underset{\alpha \to 1}{=} k\frac{V_1^{N(1-q)}-1}{1-q} + k\frac{e^{-(1-q)LN}-1}{1-q}V_1^{N(1-q)} \underset{V_1 \to +\infty}{=} \frac{k}{q-1}. \tag{20}$$

Obviously in this case the total entropy of the self-gravitating system also tends to a finite value.

On the whole, no matter in energy transfer process or in mass transfer process, as long as the nonextensive parameter is less than unity the total entropy tends to infinity, the same as the situation in classical statistics [2]; on the other hand, when nonextensive parameter is more than unity, the total entropy tends to a finite value, which is greatly different from the result of classical statistics.

In the classical statistics, the unbounded Boltzmann entropy leads to the maximum missing of entropy, which generally results in the negative heat capacity and thermodynamic instability, being thought to ubiquitously exist in the astrophysical systems. However, the ubiquitous existence of thermodynamic instability in the astrophysics has been challenged by the observations. For example, the negative heat capacity can't be directly observed and the thermodynamic remainders (black holes with small mass in orders of planets) as the results of instability can't be found everywhere in the universe. This seems to suggest that we should try to accept other ideas such as that the entropy of astrophysical systems might be bounded. Obviously, this possibility can be discussed in the realm of nonextensive statistical mechanics.

If we confine the value of nonextensive parameter in the range $q>1$, the needed conclusion immediately arises. Once the Tsallis entropy of a given self-gravitating system is bounded, the system possesses in principle a global maximum of entropy. Therefore it is possible for the self-gravitating systems to evolve into thermodynamically stable states. This is consistent with the astronomical observations: most astrophysical systems exhibit thermodynamic stability within relatively shorter time scales. In addition, just as mentioned at the beginning of this letter, the test with astrophysical data has showed that the radical velocity [15] and rotational velocity [16] of different types of stars obey the similar nonextensive power law distribution, with typical values of $q$, always more than unity.

In next section we would confirm the above value range $q>1$ in the domain of physical kinetics of self-gravitating systems.

## 4. The range of $q$ value in self-gravitating systems

Before the beginning, we firstly analyze the classical expression of Boltzmann entropy, which is the limit form of Tsallis entropy (1) in the parameter limit $q \rightarrow 1$, namely,

$$S_B = -k \int F \ln F d^{3N} \mathbf{v} d^{3N} \mathbf{r}. \tag{21}$$

The relationship between Gibbs function and distribution function of single particle can be constructed as follows,

$$F = \left(\frac{f}{N}\right)^N. \tag{22}$$

Substituting this relationship into (21), the Boltzmann entropy is transformed into

$$S_B = -Nk \int \left(\frac{f}{N}\right)^N \ln \frac{f}{N} d^{3N}\mathbf{v} d^{3N}\mathbf{r} = -k \int f \ln \frac{f}{N} d^3 \mathbf{v} d^3 \mathbf{r}. \tag{23}$$

Similar to the relation between (21) and (23), in the framework of nonextensive kinetics, we can define the Tsallis entropy as

$$S_q = -k \int f^q \ln_q f d^3 \mathbf{v} d^3 \mathbf{r}. \tag{24}$$

The distribution function $f$ in above equation satisfies the generalized Boltzmann equation [9] in the nonextensive kinetics, which of course appropriate for the self-gravitating system, namely,

$$\frac{\partial f}{\partial t} + \mathbf{v} \cdot \frac{\partial f}{\partial \mathbf{r}} - \nabla \varphi \cdot \frac{\partial f}{\partial \mathbf{v}} = C_q(f), \tag{25}$$

where $C_q$ denotes the $q$-collision term, the quantity $\varphi$ is the gravitational potential. The system described by the above generalized Boltzmann equation has $N$ hard-sphere particles of mass $m$ and diameter $s$. The $q$-collision term in (25) can be expressed as [17]

$$C_q(f) = \frac{s^2}{2} \int |\mathbf{U} \cdot \mathbf{e}| R_q d\Omega d^3 \mathbf{v}_1, \tag{26}$$

where $d^3 \mathbf{v}_1$ stands for the volume element in the velocity space, the quantity $\mathbf{U}$ is the relative velocity before collision, the $\mathbf{e}$ denotes an arbitrary unit vector, and the $s^2 d\Omega$ is the area of collision cross section.

The quantity $R_q$ is the difference of correlated distribution functions, which are assumed to satisfy the generalized form of molecular chaos hypothesis [17]. Its form is supposed to be

$$R_q(f, f') = e_q(f'^{(q-1)} \ln_q f' + f_1'^{(q-1)} \ln_q f_1') - e_q(f^{(q-1)} \ln_q f + f_1^{(q-1)} \ln_q f_1), \tag{27}$$

where the q-logarithmic function is defined as

$$\ln_q f = \frac{f^{1-q}-1}{1-q}, \tag{28}$$

and the q-exponential function is defined as [18]

$$e_q(f) = [1+(1-q)f]^{\frac{1}{1-q}}. \tag{29}$$

When the system arrives at the detailed balance, the $q$-collision term vanishes. According to the $q$-H theorem [17], the collision invariants satisfy the equality,

$$f'^{(q-1)} \ln_q f' + f_1'^{(q-1)} \ln_q f_1' = f^{(q-1)} \ln_q f + f_1^{(q-1)} \ln_q f_1. \tag{30}$$

The quantity $f'$ is the distribution function in the inverse collision process. According to the above equality, the collision invariant takes the form as follows

$$f^{q-1} \ln_q f = \frac{f^{q-1}-1}{q-1} = \gamma(\mathbf{r}) \frac{1}{2} mv^2. \tag{31}$$

Adopting this definition

$$\gamma(\mathbf{r}) = -\frac{1}{kT(\mathbf{r})}, \tag{32}$$

the incomplete distribution function is derived from (31), i.e.,

$$f(\mathbf{r},\mathbf{v}) = n(\mathbf{r})B_q \left(\frac{m}{2\pi kT(\mathbf{r})}\right)^{\frac{3}{2}} [1-(q-1)\frac{mv^2}{2kT(\mathbf{r})}]^{\frac{1}{q-1}}. \tag{33}$$

The normalized constant is given by

$$\begin{cases} B_q = \frac{1}{4}(q-1)^{\frac{3}{2}}(3q-1)\Gamma(\frac{1}{2}+\frac{1}{q-1})/\Gamma(\frac{1}{q-1}), & q>1 \\ B_q = (1-q)^{\frac{3}{2}}\Gamma(\frac{1}{1-q})/\Gamma(-\frac{3}{2}+\frac{1}{1-q}), & q\leq 1 \end{cases}. \tag{34}$$

Substituting the equation (33) into the collisionless steady Boltzmann equation, namely,

$$\mathbf{v} \cdot \frac{\partial f}{\partial \mathbf{r}} - \nabla\varphi \cdot \frac{\partial f}{\partial \mathbf{v}} = 0, \tag{35}$$

after some calculations, an important relation can be deduced [9]

$$k\nabla T + (q-1)m\nabla\varphi = 0. \tag{36}$$

This expression holds in self-gravitating systems at the hydrostatic equilibrium, which can be used to judge the value range of parameter $q$. In such systems, the long range gravity attracts all the particles together to form an obvious center whose temperature is the highest and gravitation potential is the deepest. This indicates that the direction of temperature gradient and the direction of potential gradient are always opposite to each other. This immediately leads to $q>1$.

On the other hand, through the maximum entropy method [19] or kinetic method [17], we can deduce the polytropic equation,

$$\rho(\mathbf{r}) = \rho(0)\left(\frac{T(\mathbf{r})}{T(0)}\right)^{n_{index}}, \tag{37}$$

where the relationship between polytropic index and nonextensive parameter is given by

$$n_{index} = \frac{3}{2} + \frac{1}{q-1}, \tag{38}$$

where the $n_{index}$ represents the polytropic index, whose physical origin is determined by the nature of nonextensive parameter. That is, in different systems it has different physical origin. In self-gravitating system, the polytropic index is related to the long range gravity and the system configuration. If we set $q>1$, the above index must be more than zero. According to (37), this means the mass density and temperature have the same change signs, that is, the temperature increase is accompanied by the increase of mass density, as long as the system is at hydrostatic equilibrium.

Therefore, it is reasonable to conclude that in most cases of astrophysical systems there is $q>1$; therefore the total Tsallis entropy of such systems is bounded. This leads to the possibility that most of these systems can evolve into thermodynamically stable states, at which the heat capacity of systems is positive [20].

Of course, for this we need to define a new concept called gravitational temperature which is a fundamental quantity of nonextensive thermodynamics. Based on this concept, a new heat capacity called gravitational heat capacity is then defined whose expression contains the nonextensive parameter $q$ [20]. The thermodynamic evolution, which always starts with a negative (gravitational) heat capacity, tends to increase the value of q, leading the sign change of the gravitational heat capacity ultimately.

## 5. Conclusions

In summary, we have researched the limit behavior of the entropy evolution in self-gravitating systems. No matter in the energy transfer process or in the mass transfer process, the entropy evolution exhibits similar character. Whether the Tsallis entropy is bounded or not is determined by the value of nonextensive parameter $q$. When $q<1$, the Tsallis entropy evolve to infinity; in contrast, when $q>1$, the entropy evolve to a finite value, showing a bounded property.

It is interesting that in the physical kinetics through the discussion to the generalized Boltzmann equation, an important relation linking temperature gradient and gravitational cancelation is deduced, see equation (36), which in the sense of hydrostatic equilibrium confirms that the nonextensive parameter appears to be more than unity, i.e., $q>1$. This value range is also verified by the positive definite of polytropic index (38), which is consistent with the observations about the positive correlation of density increase and temperature increase in self-gravitating systems.

Therefore, in general case the Tsallis entropy of self-gravitating systems is bounded. It is easily found that the bounded property of entropy is closely related to its $q$-logarithm definition (1); therefore the boundedness is directly derived from the nonextensivity of the entropy itself. The boundedness of Tsallis entropy indicates the existence of its global maximum.

In the realm of the nonextensive gravitational thermodynamics, as the core concept, the gravitational temperature is proposed, on basis of which the gravitational heat capacity is defined. This new heat capacity governs the thermodynamic evolution of self-gravitating system [20]. The evolution process always begins with a negative sign of gravitational heat capacity, which tends to make its sign become positive through the increase of parameter $q$. When the gravitational heat capacity of whole systems becomes positive, it is possible for these systems to evolve thermodynamically stable states.

**Acknowledgements**

This work is supported by the National Natural Science Foundation of China under Grant No.11405092, and also Grant No.11175128.